\documentclass[floatfix,superscriptaddress,showpacs,twocolumn,amssymb,amsfonts,amsmath,prl,aps]{revtex4}

\usepackage{graphicx,epsfig}

\newcommand{\dC}{$^{\circ}$C}

\begin{document}
\title{Faraday Instability in a Surface-Frozen Liquid}

\author{P.~Huber}
\email[E-mail: ]{p.huber@physik.uni-saarland.de}
\affiliation{Technische Physik, Universit\"at des Saarlandes,
66041 Saarbr\"ucken, Germany}

\author{V.P.~Soprunyuk}
\affiliation{Technische Physik, Universit\"at des Saarlandes,
66041 Saarbr\"ucken, Germany}

\author{J.P.~Embs}
\affiliation{Technische Physik, Universit\"at des Saarlandes,
66041 Saarbr\"ucken, Germany}

\author{C.~Wagner}
\affiliation{Experimentalphysik, Universit\"at des Saarlandes,
66041 Saarbr\"ucken, Germany}

\author{M.~Deutsch}
\affiliation{Physics Department, Bar-Ilan University, Ramat-Gan
52900, Israel}

\author{S.~Kumar}
\affiliation{Department of Chemical Engineering and Materials
Science, University of Minnesota, Minneapolis, MN 55455, U.S.A.}

\date{\today}

\begin{abstract}
Faraday surface instability measurements of the critical
acceleration, $a_{\rm c}$, and wavenumber, $k_{\rm c}$, for
standing surface waves on a tetracosanol (C$_{24}$H$_{50}$) melt
exhibit abrupt changes at $T_s=54~^{\circ}$C, $\sim$4~$^{\circ}$C
above the bulk freezing temperature. The measured variations of
$a_{\rm c}$\ and $k_{\rm c}$\ vs. temperature and driving
frequency are accounted for quantitatively by a hydrodynamic
model, revealing a change from a \textit{free}-slip surface flow,
generic for a \textit{free} liquid surface ($T> T_s$), to a
surface-pinned, no-slip flow, characteristic of a flow near a
wetted solid wall ($T < T_s$). The change at $T_s$ is traced to
the onset of surface freezing, where the steep velocity gradient
in the surface-pinned flow significantly increases the viscous
dissipation near the surface.
\end{abstract}

\pacs{68.10.-m, 47.20.Ma, 61.25.Em, 64.70.Dv}

\maketitle

Spatial confinement of a liquid often changes its properties
markedly. For example, superheating above the equilibrium melting
temperature \cite{conf} and order quenching upon freezing
\cite{pores} were observed under confinement only. In particular,
flow under confinement is important for processes ranging from
tribology to protein folding to transport through ion channels in
cell membranes \cite{flow}. A transition from a liquid-like to a
granular-solid-like shear response was observed at nano-scale
confinements \cite{rheo}. The surface freezing (SF) effect
\cite{SFDisc1992}, where a solid monolayer forms at the surface of
a pure normal-alkane (C$_n$H$_{2n+2}$) melt, provides a unique
system for studying semi-confined flow at a solid-liquid
interface. The abrupt onset of SF at $T_s$ allows one to switch on
(and off) the solid phase by a small temperature variation.
Understanding such interfaces would also elucidate the role of
flow in nucleation and growth processes of crystals from melts,
which are dominated by such interfaces \cite{nuc}.\\
\begin{figure}[htbp]
\epsfig{file=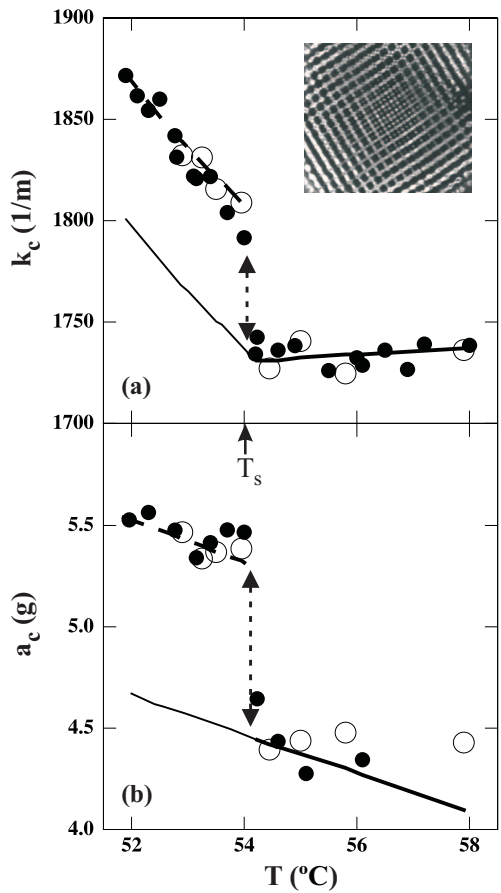, angle=0, width=0.75\columnwidth}
\caption{\label{fig1} Faraday instability parameters versus
temperature for a fixed driving frequency $f=150$ Hz. Open
(closed) symbols represent data measured on cooling (heating). (a)
Critical wavenumber, $k_{\rm c}$, (b) Critical acceleration,
$a_{\rm c}$. Inset to (a): Standing wave pattern at the surface at
$T=54$\dC. Solid and dashed lines represent the KM-model
calculations assuming free-slip and no-slip boundary conditions,
resp. Arrows mark the discontinuities in the Faraday parameters.}
\end{figure}
\indent SF occurs in melts of several chain molecules (alkenes,
alcohols, semi-fluorinated alkanes, diols, C$_i$E$_j$)
\cite{OSSF}, and at both liquid/liquid \cite{Lei2004} and
liquid/solid \cite{Merkl1997} interfaces. Related surface ordering
effects were observed in melts of polymers comprising alkyl chains
in the backbone or as side chains \cite{Gautam2002}, in liquid
alloys \cite{liqmet}, and in several liquid crystals
\cite{liqcryst}. While the structural and thermodynamic aspects of
SF have been studied in great detail \cite{Lang2004}, the
influence of surface ordering on macroscopic near-surface flows
has received little attention to date in this sizable,
technologically-important class of materials.\\
\indent To study this issue, we employ the Faraday instability,
which forms standing wave patterns (SWP) at the free surface of a
vertically-vibrated liquid \cite{Faraday1831}.  By virtue of its
simplicity, this instability is outstanding among pattern-forming
systems and a detailed theoretical description has been achieved.
The SWP formed depend sensitively on, and allow a detailed study
of, the changes in the surface hydrodynamics upon SF.  As this
study of SF demonstrates, the instability can therefore be
employed to explore physical processes which are difficult to
access by other, classical means. It also exemplifies the more
general class of parametric instabilities, which are of interest
in a broad range of areas including granular media, plasmas,
nonlinear optics, reaction-diffusion systems, and condensed-matter
physics \cite{FIGeneral}.\\
\indent The sample used is n-tetracosane, C$_{24}$H$_{50}$. With
$T_s = 54$~\dC\ and a bulk freezing temperature $T_b = 50$~\dC,\
it exhibits the largest temperature range of SF of all pure
n-alkanes \cite{SFDisc1992}. Our circular, 18-cm diameter, sample
cell was filled with 300 g of C$_{24}$H$_{50}$ (Aldrich, 99.9\%\
pure) to a height of $h=1.5$~cm above the cell bottom.  The cell
is immersed in a thermostatted water bath yielding good
temperature stability ($\pm0.05$~\dC) and minimizing temperature
gradients between the surface and bulk. The bath is vibrated
sinusoidally in the vertical direction at a driving frequency $f$
by an electromechanical shaker. The SWP are tracked and recorded
using a CCD camera and stroboscopic illumination
\cite{Mueller1997}.\\
\indent Two characteristic quantities were measured: the critical
acceleration, $a_{\rm c}$, required for destabilizing the flat
surface to form the SWP, and the critical wavenumber, $k_{\rm c}$,
which defines the spatial periodicity of the SWP at the onset of
the instability. Square SWP (inset of Fig. \ref{fig1}(a)) were
found for all $T$ and $f$ investigated. Fourier transforms were
used to determine $k_{\rm c}$ from each pattern. The
$T$-dependence of $k_{\rm c}$ and $a_{\rm c}$ at fixed $f=150$ Hz
is shown in Fig. \ref{fig1}. Upon cooling from 58~\dC\ to 54~\dC,
$k_{\rm c}$ remains roughly constant. At 54~\dC, however, a
$\sim$60~m$^{-1}$ jump is observed in $k_{\rm c}$. Below
$T=54$~\dC, $k_{\rm c}$ increases linearly with decreasing $T$. No
cooling/heating hysteresis is observed. Similar behavior is seen
for $a_{\rm c}$: it is roughly constant for 54\dC~$<T<$~58\dC,\ at
54~\dC\  it jumps by $\sim$20\%, and it is roughly constant for
$T<54$~\dC. As the discontinuities in both quantities occur right
at the onset temperature $T_s=54$~\dC\ reported for SF in
C$_{24}$H$_{50}$, it is reasonable to assign these macroscopic
effects to the onset of SF, a conclusion supported by the analysis
below.

\begin{figure}[htbp]
\epsfig{file=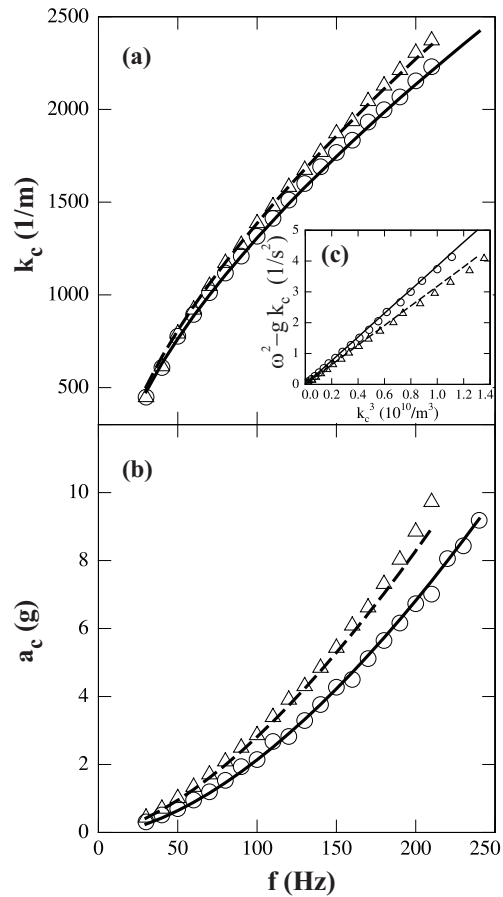, angle=0, width=0.75\columnwidth}
\caption{\label{fig2} Faraday instability parameters measured at
52~\dC~(triangles; surface-frozen state) and 58~\dC~(circles;
non-surface-frozen state) vs. the driving frequency, $f$: (a)
Critical wavelength, $k_{\rm c}$, (b) Critical acceleration,
$a_{\rm c}$. (c) Dispersion relation plot, discussed in the text.
The solid and dashed lines are calculated from the KM-model
assuming free-slip and no-slip boundary conditions, respectively.}
\end{figure}

The measured fixed-$T$\ $f$-dependence of $k_{\rm c}$ and $a_{\rm
c}$, shown in Fig. \ref{fig2}, is monotonically increasing, as
is typical for viscous Newtonian liquids \cite{Kumar1998}, both
above and below $T_s$. Both quantities are larger for the SF phase
than for the liquid surface phase over the full $f$-range
investigated, in agreement with the fixed-$f$, $T$-dependent
measurements shown in Fig. \ref{fig1}.

\begin{figure}[htbp]
\epsfig{file=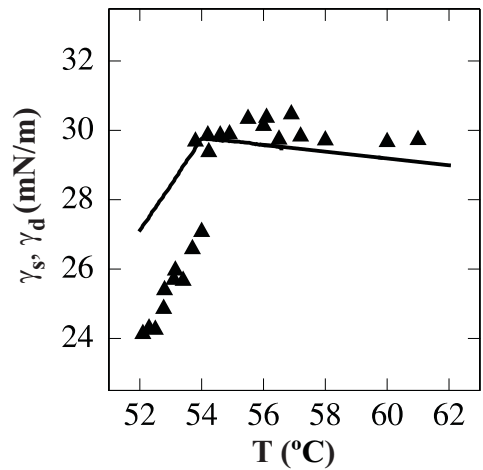, angle=0, width=0.72\columnwidth}
\caption{\label{fig3} Static surface tension $\gamma_{\rm s}$
measured by the Wilhelmy plate method (line) and dynamic surface
tension $\gamma_{\rm d}$ (solid triangles) as determined from the
surface waves' dispersion relation for a fixed $f=150$ Hz.}
\end{figure}

To account for these observations, we turn to surface
hydrodynamics. The validity of the dispersion relation of surface
waves on inviscid liquids for accurately predicting the Faraday
instability dispersion relation ($k_{\rm c}$ vs. $f$) of the SWP
has been experimentally demonstrated \cite{Benjamin1979} for
liquids of viscosities $\eta \gtrsim 100\eta_{\rm water}$. Thus,
it should also hold for liquid C$_{24}$H$_{50}$, where $\eta_{\rm
C_{24}H_{50}} \approx 4 \eta_{\rm water}$. For a liquid layer of
depth $h$ and infinite horizontal extent, this dispersion relation
is $\omega^2=(gk_{\rm c}+\gamma/\rho k_{\rm c}^3) \tanh (k_{\rm c}
h)$, where $g=9.81$ ms$^{-2}$, $\gamma$, $\rho$, and $\omega$ are
the acceleration of gravity, surface tension, density, and angular
frequency of the wave, respectively \cite{Benjamin1954}. The
observed SWP in all our experiments exhibit $\omega= 2\pi f /2$,
conforming to the common subharmonic response of the Faraday
instability \cite{Kumar1998}. Since $\tanh \left ( k_{\rm c} h
\right ) \approx 1$ in all our experiments, a simple linear
dependence of $\omega^2-gk_{\rm c}$ on $k_{\rm c}^3$ is expected,
the slope of which is $\gamma/\rho$. This relation is indeed found
both above and below $T_s$ (Fig. \ref{fig2}(c)), indicating that
in both the frozen and the non-frozen surface phases the simple
dispersion relation is obeyed. The larger slope found at
$T=58$~\dC\ as compared to that at $T=52$~\dC\ indicates that
$\gamma(T=58~^{\circ}{\rm C})>\gamma(T=52~^{\circ}{\rm C})$, since
$\rho$ does not change upon SF. This is consistent with the
standard \textit{static} surface tension curve $\gamma_{\rm
s}(T)$, measured using the Wilhelmy plate method, and shown as a
solid line in Fig. \ref{fig3}. The onset of SF at $T_s=54$~\dC\ is
clearly manifested in this curve by the abrupt change in the slope
from a negative to a positive value, due to the drop in the
surface entropy upon SF \cite{SFDisc1992}. Comparing this curve
with the \textit{dynamic} surface tension $\gamma_{\rm d}(T)$,
extracted from our measurements using the dispersion relation
above (Fig. \ref{fig3}, triangles), reveals a reasonable agreement
for the non-frozen surface phase ($T>T_s$). In the SF phase
($T<T_s$), however, a $\sim3$~mN/m shift of $\gamma_{\rm d}(T)$
below $\gamma_{\rm s}(T)$ is observed, although the linear
$T$-dependence and the slope remain the same.\\
\indent To account for this downshift, we recall that the presence
of a surfactant monolayer on a liquid surface has been
demonstrated to change the dispersion relation of surface waves
and to increase the damping \cite{Levich1962}. We have therefore
employed a model recently introduced by Kumar and Matar for the
Faraday instability in a liquid covered by an insoluble surfactant
layer (KM-model) \cite{Kumar2004}. Without an overlayer, the shear
stress at the liquid surface vanishes, yielding a free-slip
boundary condition for liquid flow. For the surfactant- or
SF-layer-covered surface, the shear stress is finite, and in the
limit of large Marangoni number (ratio of surface-tension-gradient
forces to viscous forces) yields a pinned, no-slip boundary
condition. We use the KM-model with the former boundary condition
for $T>T_{\rm s}$, and with the latter boundary condition for
$T<T_{\rm s}$. Given the crystalline, solid structure of the
surface-frozen monolayer, such an assumption appears justified.
Using the measured static surface tension $\gamma_{\rm s}(T)$, the
dynamic viscosity $\eta(T)$ \cite{Beiner2004}, and the
aforementioned change in the boundary condition at $T_{\rm s}$,
the KM-model yields for fixed $f=150$ Hz the lines shown in Fig.
\ref{fig1} (a) and (b). The good agreement with the measured
values is evident. In particular, the change of the hydrodynamic
boundary condition at $T_{\rm s}$ results not only in a jump of
$\sim$20\%\ in $a_{\rm c}$, but also in a similar jump of
$\sim$3\% in $k_{\rm c}$, both in good agreement with the
experimental observation. Moreover, the frequency dependent
calculations, shown as lines in Fig. \ref{fig2}, demonstrate that
the KM-model accounts accurately also for the observed frequency
dependence of $k_{\rm c}$ and $a_{\rm c}$ over the full dynamic
range investigated, and for the abrupt increase in these
quantities upon SF.\\
\indent The increase in $a_{\rm c}$ upon SF reflects an increase
in the damping of the surface excitations, assignable, in turn, to
the large velocity gradient near the surface imposed by the
pinning of the flow at the surface. The increased damping leads to
a viscous detuning of the inviscid dispersion relation, and thus
to a jump in the spatial periodicity of the waves. It is
interesting to note that the change in the boundary condition is
equivalent to an infinite surface elasticity: the surface in our
model can still deform in the vertical direction, but the
deformations are such that the surface neither contracts nor
expands.\\
\indent One of the seminal papers on SF in n-alkanes reports
light-scattering measurements on thermally-excited capillary waves
above and below $T_{\rm s}$ \cite{Hughes1993}. The surface
elongations due to capillary waves are microscopic, typically of
the order of nanometers \cite{Braslau1985}. However, they are
still governed by the same dispersion relation as the macroscopic,
mm-scale surface waves of the Faraday instability
\cite{Langevin1992}. It therefore seems puzzling that for
capillary waves a decreased damping was observed upon SF, rather
than the increased damping found here. We believe the different
behavior arises due to the different magnitudes of the surface
elongations in the two cases, and consequently the different
nature of the excited motion. The Faraday instability excites
macroscopic waves which produce real hydrodynamic flow in the
proximity of the surface. Thus, the no-slip boundary condition in
the surface-frozen phase induces an increased velocity gradient,
and consequently viscous dissipation across macroscopic
surface-normal distances ensues. In contrast, for the
nanometer-sized amplitudes of the thermally-excited capillary
waves, almost no macroscopic hydrodynamic flow occurs. The
postulated conformational changes in the molecular shape at
$T_{\rm s}$, from a flexible, end-distorted shape to a rigid
extended one \cite{conformal}, may even enhance the molecular
mobility, particularly along their long, surface-normal, axis.
This, in turn, should reduce the damping for these microscopic
excitations. Indeed, the increase in surface-normal molecular
mobility, which increases the surface entropy
\cite{Tkachenko1996}, and the molecular conformation changes,
which reduce the internal energy \cite{conformal}, are among the
strongest candidates proposed for explaining the occurrence of the
SF effect.\\
\indent The SF-induced changes in the boundary condition,
demonstrated here to alter significantly the macroscopic flow
characteristics near the surface for one particular alkane, should
be of importance for all materials exhibiting surface or
interfacial freezing, particularly when the flow geometry involves
large surface areas. One example is microfluidic devices
\cite{Stone2004}. Another example is foams, where the thinning and
bursting rates of the bubbles are determined by the viscous
drainage flow within the liquid films constituting the bubble
walls \cite{Debregeas1998}. Thus, the present results explain also
the hydrodynamics underlying the increased lifetime of alkane
bubbles, reported as a macroscopic manifestation of SF
\cite{Gang1998}. The increase in the effective viscous drag on the
drainage flow, caused by the no-slip boundary condition in the SF
phase, reduces the drainage rate, and consequently the
wall-thinning and bursting rates of the bubbles. This leads to the
observed increase in the bubbles' lifetime upon the onset of SF.\\
\indent It has recently been shown that the Faraday instability
can be employed to explore rheological behavior of bulk liquids
\cite{Raynal1999}, especially when it is difficult to use other
techniques (e.g., near a liquid-vapor critical point
\cite{Fauve1992}). We demonstrated here that this method allows
visualizing and examining on a \textit{macroscopic} scale
rheological aspects of \textit{microscopic} structured surfaces,
in particular the interesting case of a change of the shear-stress
boundary condition at a liquid's surface from the generic,
free-slip one to a no-slip one, characteristic of a wetted solid
wall \cite{noslip}. We hope that this study will stimulate further
experiments focusing on how the surface hydrodynamics is affected
by microscopic modifications of the structure of surfaces or
interfaces, e.g., experiments on wetting transitions which have
been proven to allow for a precise control of liquid surface
microstructure as a function of temperature \cite{wetting}. This
method could also extend the wavelength and frequency ranges of
established semi-microscopic and microscopic techniques like light
scattering \cite{Langevin1992} and surface x-ray photon
correlation spectroscopy \cite{Gutt2003} towards macroscopic
hydrodynamic length scales. More generally, the present method
could be applied to other parametric instabilities to provide
deeper insight into those areas of science where such
instabilities occur \cite{FIGeneral}.\\
\indent We thank K. Knorr and M. L\"ucke for helpful discussions
and acknowledge support by the DFG (SFB 277).


\begin{references}

\bibitem{conf} F. Banhart, E. Hernandez, and M. Terrones, Phys. Rev.
Lett. \textbf{90}, 185502 (2003); L. Zhang \textit{et al.}, ibid.
\textbf{85}, 1484 (2000).

\bibitem{pores} P. Huber \textit{et al.}, Europhys. Lett. \textbf{65}, 351
(2004).

\bibitem{flow} M. Urbakh \textit{et al.}, Nature \textbf{430}, 525 (2004);
J.M. Drake and J. Klafter, Physics Today \textbf{43}, 46 (1990).

\bibitem{rheo} U. Raviv, P. Laurat, and J. Klein, Nature
\textbf{413}, 51 (2001).

\bibitem{SFDisc1992} J.C. Earnshaw and C.J. Hughes, Phys. Rev. A
\textbf{46}, R4494 (1992); X.Z. Wu \textit{et al.}, Phys. Rev.
Lett. \textbf{70}, 958 (1993); E.B. Sirota \textit{et al.} ibid.
\textbf{79}, 531 (1997) ;X.Z. Wu \textit{et al.} Science,
\textbf{261},1018 (1993); B.M. Ocko \textit{et al.}, Phys. Rev. E
\textbf{55}, 3164 (1997).

\bibitem{nuc} S. Butler and P. Harrowell, Nature \textbf{415},
1008 (2002); X.Y Liu \textit{et al.}, ibid. \textbf{374}, 342
(1995).

\bibitem{OSSF} H. Gang \textit{et al.} J. Phys. Chem. B
\textbf{102}, 2754 (1998); M. Deutsch \textit{et al.} Europhys.
Lett. \textbf{30}, 283 (1995);
O. Gang \textit{et al.}, ibid. \textbf{49}, 761 (2000).

\bibitem{Lei2004} Q. Lei and C.D. Bain, Phys. Rev. Lett.
\textbf{92}, 176103 (2004).

\bibitem{Merkl1997} C. Merkl, T. Pfohl, and H. Riegler, Phys. Rev. Lett. \textbf{79} 4625 (1997); U. Volkmann et. al. J. Chem. Phys. \textbf{116}, 2107
(2002).

\bibitem{Gautam2002} K.S. Gautam and A. Dhinojwala, Phys. Rev. Lett. \textbf{88}, 145501 (2002); K.S. Gautam \textit{et al.}, ibid. \textbf{90}, 215501 (2003).

\bibitem{liqmet} A. Turchanin, D. Nattland, and W. Freyland, Chem. Phys. Lett. \textbf{337}, 5 (2001).

\bibitem{liqcryst} J. Als-Nielsen, F. Christensen, and P.S. Pershan, Phys. Rev. Lett. \textbf{48}, 1107 (1982); B.M. Ocko \textit{et al.}, ibid. \textbf{57}, 94 (1986) ; X.F.
Han et al., ibid. \textbf{91}, 045501 (2003).

\bibitem{Lang2004} P. Lang, J. Phys.: Condens. Matter \textbf{16},
R699 (2004).

\bibitem{Faraday1831} M. Faraday, Phil. Trans. R. Soc. London
\textbf{52}, 319 (1831).

\bibitem{FIGeneral} J. R. de Bruyn \textit{et al.}, Phys. Rev. Lett. \textbf{81}, 1421 (1998);
P. K. Shukla, ibid. \textbf{84}, 5328 (2000); C. J. McKinstrie
\textit{et al.}, Optics Express \textbf{11}, 2619 (2003); V.
Petrov, Qi Ouyang, and Harry L. Swinney, Nature \textbf{388}, 655
(1997); G. M. Genkin, Phys. Rev. A \textbf{63}, 025602 (2001).


\bibitem{Mueller1997} C. Wagner, H.W. M\"uller, and K. Knorr, Phys. Rev. E \textbf{68}, 066204 (2003).

\bibitem{Kumar1998} K. Kumar and L. S. Tuckerman, J. Fluid Mech.
\textbf{279}, 49 (1994).

\bibitem{Benjamin1979} T.B. Benjamin and J.C. Scott, J. Fluid Mech.
\textbf{92}, 241 (1979); W.S. Edwards and S. Fauve, ibid. \textbf{278}, 123 (1994).

\bibitem{Benjamin1954} T.B. Benjamin and F. Ursell, Proc. R. Soc.
London A \textbf{225}, 505 (1954).

\bibitem{Levich1962} V.G. Levich, Physicochemical
Hydrodynamics (Prentice-Hall, Englewood Cliffs, NJ, 1962) ; J.
Miles and D. Henderson, Ann. Rev. Fluid Mech. \textbf{22}, 143
(1990);  J.A. Nicolas, J.M. Vega, and J. Fluid Mech. \textbf{410},
367 (2000).

\bibitem{Kumar2004} S. Kumar and O.K. Matar, Phys. Fluids
\textbf{16}, 39 (2004).

\bibitem{Beiner2004} $\eta$ increases slightly from $4.5$ mPa s to $5$ mPa s upon cooling
from $65$\dC~to $52$\dC (M. Beiner,~U. Halle-Wittenberg).

\bibitem{Hughes1993} C.J. Hughes and J. C. Earnshaw, Phys. Rev. E
\textbf{47}, 3485 (1993).

\bibitem{Braslau1985} A. Braslau \textit{et al.} Phys. Rev. Lett. \textbf{54}, 114 (1985);
B.M. Ocko \textit{et al.} ibid. \textbf{72}, 242 (1994).

\bibitem{Langevin1992} Light Scattering by Liquid Surfaces and
Complem. Techn., ed. by D. Langevin (Marcel Dekker, NY, 1992).

\bibitem{conformal} A. J. Colussi, M.R. Hoffmann, and Y.C. Tang, Langmuir \textbf{16}, 5213
(2000).

\bibitem{Tkachenko1996} A.V. Tkachenko and Y. Rabin, Phys. Rev.
Lett. \textbf{76}, 2527 (1996).

\bibitem{Stone2004} H.A. Stone, A.D. Stroock, and A. Ajdari, Ann. Rev. Fluid Mech. \textbf{36}, 381 (2004).
\bibitem{Debregeas1998} G. Debregeas, P.G. de Gennes, and F. Brochard-Wyart, Science \textbf{279}, 1704 (1998).

\bibitem{Gang1998} H. Gang \textit{et al.} Europhys. Lett \textbf{43},
314 (1998).

\bibitem{Raynal1999} F. Raynal \textit{et al.}, Eur. Phys. J. B \textbf{9}, 175 (1999).

\bibitem{Fauve1992} S. Fauve \textit{et al.}, Phys. Rev. Lett. \textbf{68}, 3160 (1992).

\bibitem{noslip} E. Lauga, M.P. Brenner, and H.A. Stone, cond-mat/0501557.
\bibitem{wetting} H. Tostmann \textit{et al.} Phys. Rev. Lett. \textbf{84}, 4385
(2000); P. Huber \textit{et al.}, ibid. \textbf{89}, 035502
(2002); D. Bonn and D. Ross, Rep. Prog. Phys. \textbf{64}, 1085
(2001).
\bibitem{Gutt2003} T. Seydel \textit{et al.}, Phys. Rev. B \textbf{63}, 073409 (2001); A. Madsen \textit{et al.}, Phys. Rev. Lett. \textbf{90}, 085701 (2003); C. Gutt \textit{et al.}, Phys. Rev. Lett. \textbf{91}, 076104
(2003).
\end{references}
\end{document}